\documentclass[conference]{IEEEtran}
\IEEEoverridecommandlockouts
\usepackage{cite}
\usepackage{amsmath,amssymb,amsfonts}

\usepackage{graphicx}
\usepackage{textcomp}
\usepackage{xcolor}
\usepackage{algorithm}
\usepackage{algpseudocode}
\usepackage{pifont}
\usepackage[caption=false]{subfig}
\def\BibTeX{{\rm B\kern-.05em{\sc i\kern-.025em b}\kern-.08em
    T\kern-.1667em\lower.7ex\hbox{E}\kern-.125emX}}
\begin{document}

\title{Histogram-based Auto Segmentation: A Novel Approach to Segmenting Integrated Circuit Structures from SEM Images}

\author{\IEEEauthorblockN{Ronald Wilson, Navid Asadizanjani, Domenic Forte and Damon L. Woodard}
\IEEEauthorblockA{\textit{Florida Institute for Cyber Security Research} \\
\textit{University of Florida}\\
Gainesville, FL, USA}
}

\maketitle
\begin{abstract}
In the Reverse Engineering and Hardware Assurance domain, a majority of the data acquisition is done through electron microscopy techniques such as Scanning Electron Microscopy (SEM). However, unlike its counterparts in optical imaging, only a limited number of techniques are available to enhance and extract information from the raw SEM images. In this paper, we introduce an algorithm to segment out Integrated Circuit (IC) structures from the SEM image. Unlike existing algorithms discussed in this paper, this algorithm is unsupervised, parameter-free and does not require prior information on the noise model or features in the target image making it effective in low quality image acquisition scenarios as well. Furthermore, the results from the application of the algorithm on various structures and layers in the IC are reported and discussed.
\end{abstract}


\section{Introduction}

Reverse engineering (RE) is the science of understanding the constituent components from a final product. It is essentially the entire engineering work flow performed in reverse. This engineering technique can be applied to a variety of products ranging all the way from aircraft to Integrated Circuits (IC). As intuition suggests, the higher the complexity in the product, the higher the hardships involved in RE the product. This is especially true with the present day ICs with several billion transistors and interconnections occupying a very small area on a silicon wafer.

Despite the obvious shortcoming of illegal duplication of proprietary technology, RE has several benefits. One of them is to have a better understanding of the structure of the final product and the effect of the physical processes used in the manufacturing of the product. For instance, RE methods were used to analyze and debug an IC \cite{c1}. Similar applications can be seen in Trojan detection \cite{c2, c23} and settling legal disputes between companies on infringement of intellectual property usage.

In the past, RE of IC was mostly done with the help of subject matter experts. Detailed optical images of the IC die were taken and components marked down by hand \cite{c3}. This was a very laborious process. However, with the incorporation of image analysis algorithms into RE, the entire process became easier and semi-automated. A simple method is described in \cite{c4}, where the image was processed using a median filter followed by correlation matching. But with the introduction of higher node technologies, optical imaging became obsolete and required imaging modalities with higher resolution such as Scanning Electron Microscopy (SEM). The approaches developed for handling issues with optical images cannot be reliably applied to SEM images. Some of these issues such as lack of understanding of noise models, variation in size of relevant features and others are discussed in the detail along with their implications for hardware trust and assurance in \cite{c13}.

\section{The Imaging Modality}

SEM images are produced by accelerating electrons towards a region of interest and observing the interactions of the electrons with the target materials. There are two main types of interactions: Secondary Electrons (SE) and Back Scattered Electrons (BSE). The quality of these two interactions depend on the constituent materials in the imaged IC and several parameters set by the operator such as:
\begin{itemize}
\item Excitation Voltage: The excitation voltage of the electrons control the depth of penetration into the sample. The higher the voltage, the higher the penetration. 
\item Magnification: This parameter helps zoom into the image. Small features can be easily imaged by adjusting this parameter. 
\item Resolution: The parameter refers to the number of pixels in the image. The higher the pixel count, the better the quality of the image. 
\item Dwelling Time: This refers to the time the scanning beam takes to measure a single pixel in the image. The higher the time spent, the better the quality of the image. 
\end{itemize}

\begin{figure}
    \centering
    \includegraphics[clip,width=0.95\columnwidth]{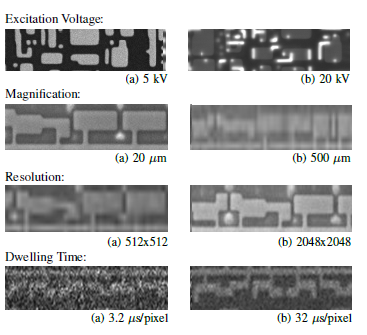}
    \caption{Affect of tuning imaging parameters in SEM \cite{c10, c13}}
    \vspace{-5mm}
    \label{fig:image_param}
\end{figure}

The affect of tuning these parameters can be seen in Figure \ref{fig:image_param}. In order to image an IC, the IC has to be depackaged and delayered. Depackaging involves the extraction of the die from its case and delayering performs removal of materials from the die at a set depth. A detailed description of the depackaging and delayering process has been described in \cite{c4, c5}. After delayering, the IC is imaged in a row raster fashion. The type of image to be selected for further processing can be based on the criteria set forth in earlier works \cite{c6}. In our case, we have chosen the SE images for all our experiments. However, this algorithm can also be applied to BSE images.

Even though the SEM produces images of considerably high resolution, it does have inherent noise that introduces artifacts in the image. All the methods proposed previously, to the best of our knowledge, overcame the inherent flaws by tuning the parameters discussed above. The effect of tuning the parameters has been studied before with higher quality images taking over 30 days to process \cite{c10}. The time frame to process images for a higher node technology would be unfeasible. This is one of the major obstacles to RE and one of the problems this paper would tackle. 

{}
   
In order to reduce the time requirements, it will be necessary to use lower quality images. This implies using images with the inherent noise corruption. There are several sources of noise in the imaging modality. Some of them are topography of the material, manufacturing defects, diffusion, damage from deprocessing the IC, atmospheric exposure during deprocessing, electron transmigration from prior usage and conductivity issues \cite{c13}. The degree of contribution from each one of these noise sources is hard to model. Furthermore, there might be other sources of noises that has not been fully understood yet. 

Segmentation methods on SEM images currently employed in the field of RE involves two distinct approaches. The first approach relies on spatial and frequency domain filtering typically followed by thresholding \cite{c4, c14, c15, c16}. The primary concern behind the usage of these methods are the assumptions on the noise model corrupting the SEM images. Applying filters based on their effectiveness on optical images is not necessarily guaranteed to carry over to SEM images. The second approach relies on machine learning including deep learning \cite{c11, c17, c18, c19, c20, c22, c23}. The methods rely on trained models which require a lot of labeled data while still producing faulty segmentation. Moreover, these features are hand-picked from image patches by the user. Due to the inherent nature of the ICs, the type and size of features are drastically different at various layers \cite{c21}. Therefore, an exhaustive set of features cannot be determined efficiently. Most of these algorithms, hence, prioritize the simple features in the metal and contact layer.

\section{Our Approach}

The proposed algorithm has only one optional parameter. This is the size of the smallest feature that would be present in the target image. For our application, this would be the cross-section for the Vertical Interconnect Access (via) in our imaging plane. However, the parameter can be set to a default kernel size of 2x2. The idea is to use the largest kernel size possible in order to speed up the segmentation. There are three main steps in the algorithm.

\subsection{Filtering image using simple merge and optimization} \label{step1}

The initial step is to extract the actual histogram of the image. In order to do that, the image needs to be filtered. We employ a median filter of the fixed kernel size and convolve it with the image in a row raster fashion using stride equal to the width of the kernel. This ensures no overlap between two consecutive kernel patches. The median value is calculated and the difference between each consecutive kernel patches is extracted. Once a row is complete, a frequency distribution of the difference between consecutive medians are taken. The differences are stored in `$\alpha$' and their corresponding frequencies in  `$\beta$'.

\begin{equation} \label{eq:1}
\tau = min \Bigg[\beta(i) - \beta(i) * \alpha(i)\Bigg]  
\end{equation}

Next, an optimization is performed using Equation \ref{eq:1} and a merge threshold `$\tau$' for the current row is calculated. All kernel blocks with median difference less than `$\tau$' are merged together and their combined median is extracted. All pixels in these kernels will have their intensity values replaced by the combined median value. This process is performed on all rows.
Once completed, the frequency distribution of pixel intensity values is extracted from the filtered image. This is the estimated histogram for the image.

\subsection{Finding local peaks in the histogram} \label{step2}
In this step, we will extract the local maxima from the estimated histogram for the image. This is performed using an accumulator generated by algorithm \ref{alg1}. The algorithm highlights the most likely local peaks in the histogram. These peaks are taken to be the different materials present in the image.   

\begin{algorithm}
\caption{To find significant peaks in histogram}
\label{alg1}
\begin{algorithmic}[1]
\State Sort intensity using corresponding frequency in decreasing order and save as $\bf{I}$
\State Set $\bf{I}$ $\leftarrow$ $\bf{I}$ - $\bf{I}$[max(frequency)]
\State Extract positive values from $\bf{I}$ and save as $\bf{F}$
\State Initialize currentIntensity to 1
\State Initialize an accumulator array of size(F)
\While{currentIntensity $<$ max(F)}
\State Find index of currentIntensity in F and save as $f$ 
\State Assign all elements from F[0] to F[f] into G
\State  Initialize flag to True
\While{flag is True}
\If{currentIntensity + 1 in G}
\State currentIntensity $\leftarrow$ currentIntensity + 1
\Else
\State currentIntensity $\leftarrow$ currentIntensity + 1
\State Set flag as False
\EndIf
\EndWhile
\For{any e $\in$ G $>$ currentIntensity}
\State accumulator[e] $\leftarrow$ accumulator[e] + 1
\EndFor     
\EndWhile
\State Element-wise multiply accumulator with corresponding indices in unsorted intensity
\State Threshold accumulator using its mean value
\State Repeat Steps from 3 to 23 for $\mid$negative values$\mid$ in I
\State Join left and right accumulators preserving order  
\end{algorithmic}
\end{algorithm}

\subsection{Finding the decision boundaries} \label{step3}

This is the final stage of the algorithm. In this step, we utilize the local peaks obtained from the previous step to decide the boundary intensities of the different materials present in the image. We have two approaches to deciding the decision boundaries: 
\begin{itemize}
\item Distance-based: The euclidean distance of the candidate intensity from the detected peaks.
\item Histogram-based: The intensity of minimum frequency between two detected peaks.
\end{itemize}
In our results, the histogram-based method is used.

\section{Discussion}

The vias, shown in Figure \ref{poly}, are the smallest feature for our samples and their size was found to be bounded between a kernel of size 4x4. Hence, we set the kernel size parameter to 3x3. 

The primary problem the paper is trying to address can also be seen in Figure \ref{poly}(c). The noise in the image is causing the polysilicon regions to merge. Unlike common applications of unsupervised segmentation in which the bounding regions does not have to be kept precise, a merge of polysilicon regions during the segmentation process can affect the entire functionality of the final RE product. Hence, it is better to slightly over-segment the polysilicon regions than under-segment it. In this way, we can avoid incorrectly merged regions.

\begin{figure}[]
\centering
\subfloat[Raw Image Histogram]{\includegraphics[clip,width=0.85\columnwidth]{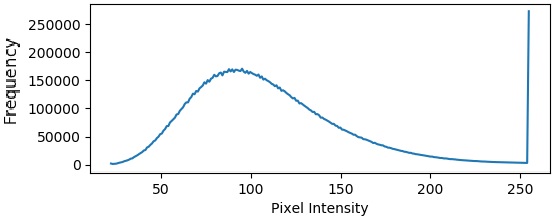}}$ $
\subfloat[Extracted Image Histogram]{\includegraphics[clip,width=0.85\columnwidth]{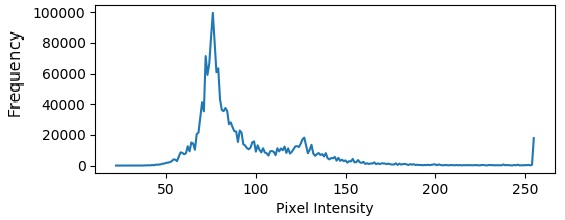}}
\caption{Histogram Correction}
\vspace{-5mm}
\label{hist_corrected}
\end{figure}

The histogram of the raw polysilicon layer from Figure \ref{poly}(a) is shown in Figure \ref{hist_corrected}(a). It can be reasoned from the histogram that there are only two possible peaks. However, it can be seen from Figure \ref{poly}(a) that there are three different materials in the image: the polysilicon structures, the vias and the silicon substrate. This ambiguity prompted the need to pre-process the image histogram using the filtering step mentioned in Section \ref{step1}. The corrected histogram can be seen in Figure \ref{hist_corrected}(b). There are three distinct peaks in the estimated histogram, thereby, resolving the ambiguity.  

Once the histogram is estimated, the next step is to calculate the decision boundaries between the different materials. This is, however, difficult due to the noise in the histogram. A simple smoothing would be effective with the risk of smoothing smaller peaks. Hence, algorithm \ref{alg1} from Section \ref{step2} was used to extract the peaks.

\begin{figure}[]
\centering
\subfloat[Raw accumulator]{\includegraphics[clip,width=0.85\columnwidth]{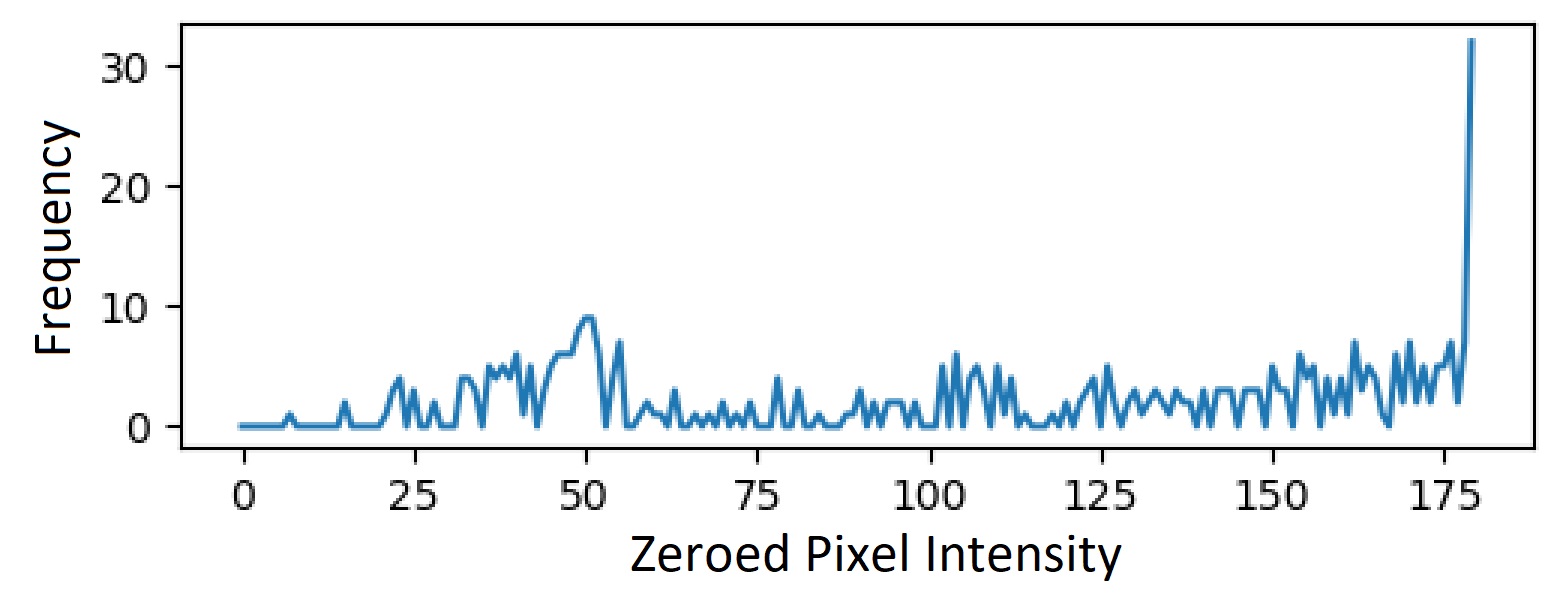}}$ $
\subfloat[Accumulator multiplied with frequency]{\includegraphics[clip,width=0.85\columnwidth]{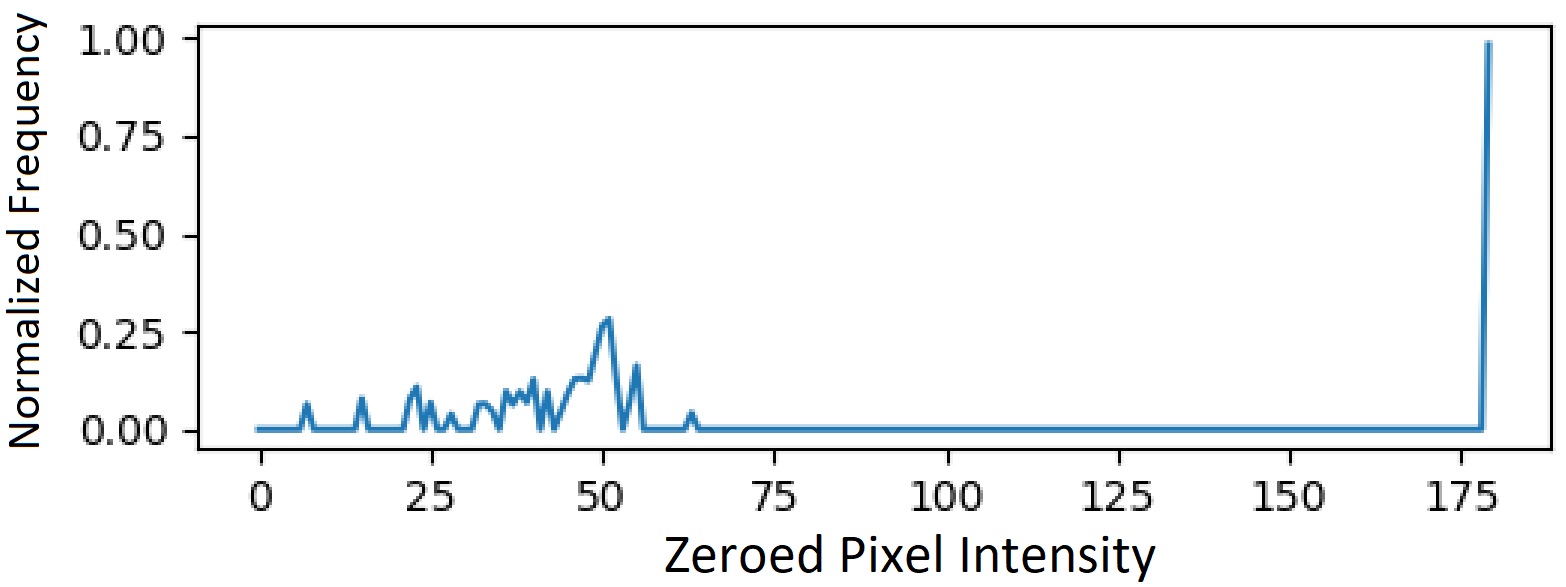}}
\caption{Accumulator [To the right direction only]}
\label{accumulator}
\end{figure}

The accumulator described in the algorithm (algorithm \ref{alg1} steps 1-21) can be seen in Figure \ref{accumulator}(a). The basic idea behind the accumulator is that peaks that are farther away from the global maxima get more votes. This is a reasonable assumption. However, the number of votes may be higher than the actual peaks itself. Hence, the accumulator is multiplied with the corresponding frequencies (algorithm \ref{alg1} step 22). This scales down the accumulator so that only points with high frequency remain and the rest are suppressed. The result of this step in shown in Figure \ref{accumulator}(b). The suppression process is enhanced further by thresholding the accumulator using the mean of all the values (algorithm \ref{alg1} step 23).   

\begin{figure}[]
\centering
\includegraphics[width=0.85\columnwidth]{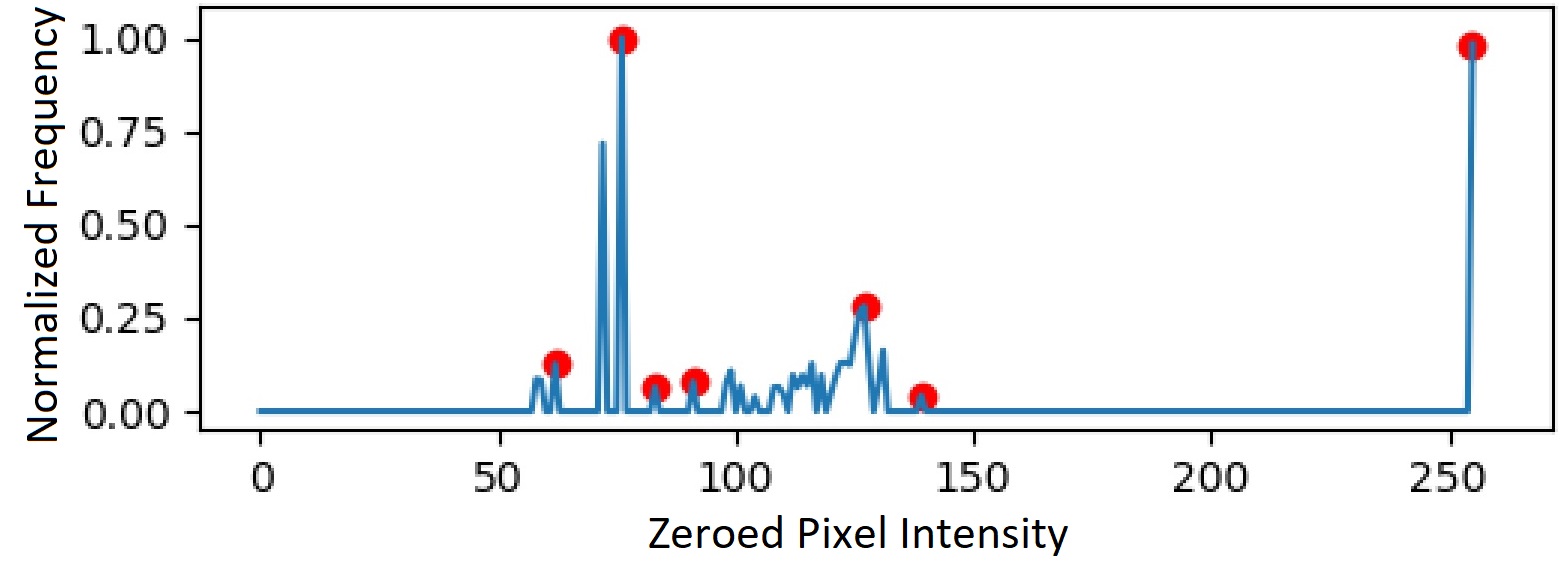}
\caption{Merged and thresholded accumulator showing selected peaks}
\vspace{-5mm}
\label{thresh_accum}
\end{figure}

Thresholding the accumulator yields the graph shown in Figure \ref{thresh_accum}. The process results in distinct peaks with discontinuities between them. The peaks that are in contact are merged into one. The peaks that remain are highlighted using red dots in the plot. Once the peaks are obtained, the histogram is divided between the peaks as described in Section \ref{step3}. The member intensities of each peak can be decided using a simple distance metric such as euclidean distance or by using the lowest frequency point between the peaks in the histogram itself. The variation in the segmentation from these two methods is very small. It stands to reason that, the better the representation of the image by the histogram, the better the results using the latter method. 

\section{Results}

\begin{table*}[h]
\caption{Segmented results from applying various filters to all layers (Threshold = 108 selected using ground truth)}
\vspace{-5mm}
\label{Results}
\begin{center}
\begin{tabular}{@{}|c|c|c|c|c|c|c|c|c|@{}}
\hline
Layer & Raw image & 
\begin{tabular}[c]{@{}c@{}}Anisotropic\\ Diffusion\end{tabular}& 
\begin{tabular}[c]{@{}c@{}}Curvature\\filter\end{tabular} & 
\begin{tabular}[c]{@{}c@{}}Gaussian\\filter\end{tabular}& 
\begin{tabular}[c]{@{}c@{}}Median\\filter\end{tabular} & \begin{tabular}[c]{@{}c@{}}HAS\\(Distance)\end{tabular} & 
\begin{tabular}[c]{@{}c@{}}HAS\\(Histogram)\end{tabular} & 
Ground Truth \\
\hline

& & & & & & & &\\
Doped &
\includegraphics[width=0.2\columnwidth]{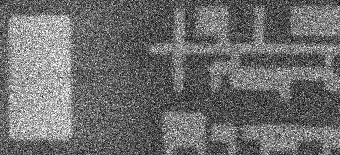} & 
\includegraphics[width=0.2\columnwidth]{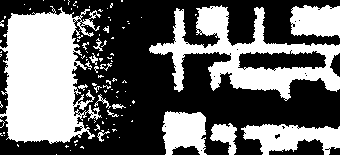} & 
\includegraphics[width=0.2\columnwidth]{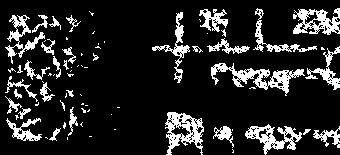} &
\includegraphics[width=0.2\columnwidth]{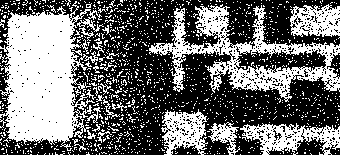} & 
\includegraphics[width=0.2\columnwidth]{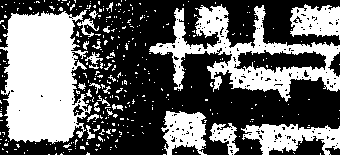} & 
\includegraphics[width=0.2\columnwidth]{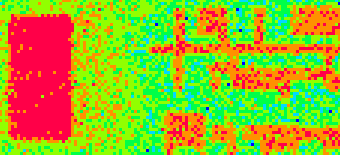} &
\includegraphics[width=0.2\columnwidth]{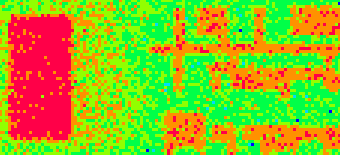} &
\includegraphics[width=0.2\columnwidth]{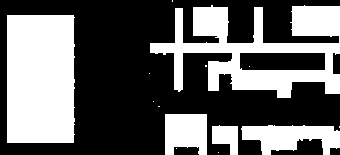}\\ 
& & & & & & & &\\
\begin{tabular}[c]{@{}c@{}}Poly\\silicon\end{tabular} &
\includegraphics[width=0.2\columnwidth]{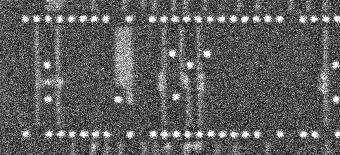} & 
\includegraphics[width=0.2\columnwidth]{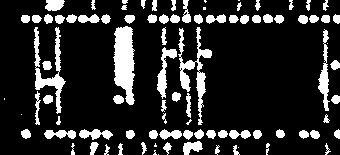} & 
\includegraphics[width=0.2\columnwidth]{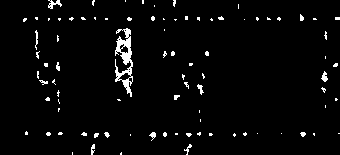} &
\includegraphics[width=0.2\columnwidth]{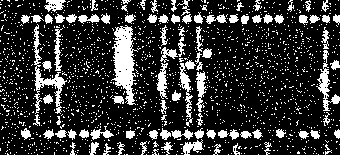} & 
\includegraphics[width=0.2\columnwidth]{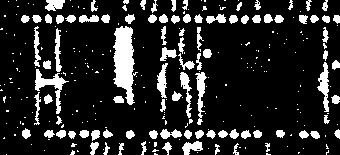} & 
\includegraphics[width=0.2\columnwidth]{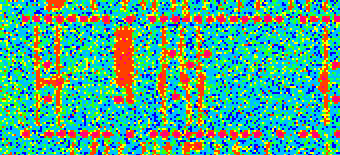} &
\includegraphics[width=0.2\columnwidth]{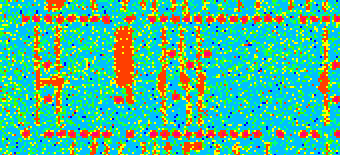} &
\includegraphics[width=0.2\columnwidth]{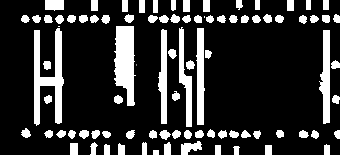}\\

Metal &
\includegraphics[width=0.2\columnwidth]{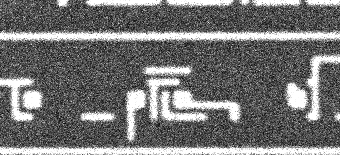} & 
\includegraphics[width=0.2\columnwidth]{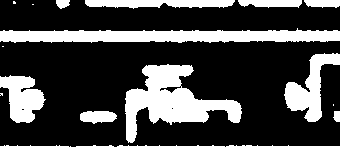} & 
\includegraphics[width=0.2\columnwidth]{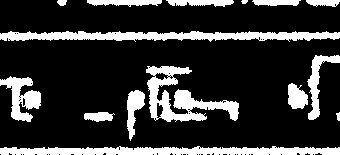} &
\includegraphics[width=0.2\columnwidth]{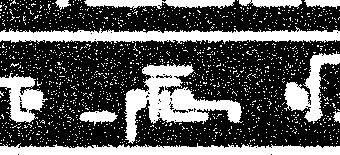} & 
\includegraphics[width=0.2\columnwidth]{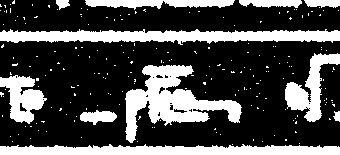} & 
\includegraphics[width=0.2\columnwidth]{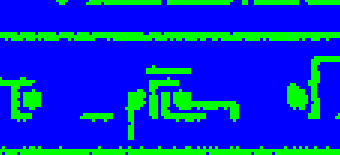} &
\includegraphics[width=0.2\columnwidth]{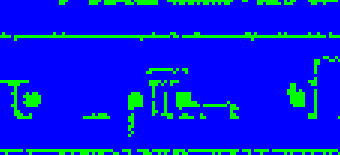} &
\includegraphics[width=0.2\columnwidth]{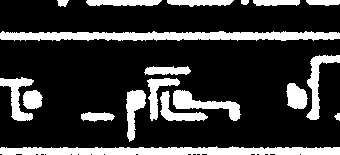}\\
\hline
\end{tabular}
\end{center}
\end{table*}

The algorithm was applied to three main types of layers found in IC: the doped region, polysilicon region and the metal interconnect layer. The source images were taken from a SmartCard IC with the parameters: 150 $\mu$m, 10 $\mu$s/pixel, 4096x4096 at 5 kV for the magnification, dwelling time, resolution and excitation voltage respectively.  

\begin{figure}
    \centering
    \includegraphics[width=0.8\columnwidth]{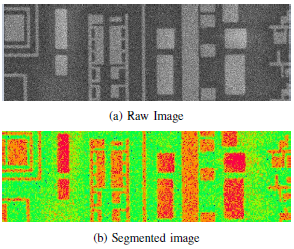}
    \caption{Doped Region [Doping(red/orange), Silicon substrate(Green)]}
    \label{dope}
    \vspace{-5mm}
\end{figure}{}

\begin{figure}
    \centering
    \includegraphics[width=0.7\columnwidth]{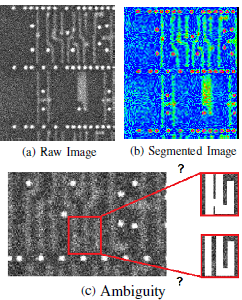}
    \caption{Polysilicon layer [Vias(red), Polysilicon(green), silicon substrate(blue)] \cite{c13}}
    \label{poly}
    \vspace{-4mm}
\end{figure}

\begin{figure}
    \centering
    \includegraphics[width=0.75\columnwidth]{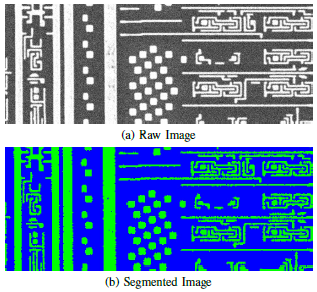}
    \caption{Metal Layer [Metal(green), silicon substrate(blue)]}
    \label{metal}
    \vspace{-6mm}
\end{figure}

The doped region and its segmented result is depicted in Figure \ref{dope}. Doped regions are usually easy to segment. However, due to the inherent noise in the image, as described earlier, the image can have variation in its intensity. This can be seen as bright regions on the lower left and right of the image. This forces the algorithm to assign them to  different peaks. But, the shape of the structure is still conserved. The polysilicon region is shown in Figure \ref{poly}(a). This is the hardest to segment. The noise in this region is much higher and any imperfections in the segmentation would cause a major setback in the RE workflow. The shape of the structures determines its functionality in the completed circuit. Even though the segmentation depicted in Figure \ref{poly}(b) is still a bit noisy, it still extracts the shape of the structure along with the vias. The noise is mostly concentrated on the silicon substrate which is inconsequential to the RE process. The metal layer and its segmented result is shown in Figure \ref{metal}. Due to the nature of the materials used, the effect of the inherent noise sources on these types of images is minimal. Hence, they are easy to segment.      

{}  

{}

\begin{figure}
    \centering
    \includegraphics[clip, width=0.8\columnwidth]{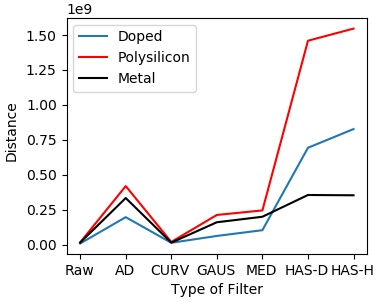}
    \caption{Performance of various filtering algorithms on SEM images. Higher value indicates better performance. [Raw: Raw image, AD: Anisotropic Diffusion, CURV: Curvature filters (Gaussian Curvature), GAUS: Gaussian smoothing, MED: Median filter, HAS-D/H: Distance/Histogram]}
    \vspace{-5mm}
    \label{fig:alg_perf}
\end{figure}

We have also performed a comprehensive comparison of all existing image analysis techniques for hardware assurance to that of our algorithm on all three layers. The results are shown in Figure \ref{fig:alg_perf} and Table \ref{Results}. The ground truth for the pixels in the images were labelled manually. Each pixel was labelled as foreground or background and their corresponding distributions were obtained. For a good quality image segmentation, the distributions between foreground and background will not overlap and, therefore, produce a large distance between them. We have used this method as a means of comparison between different techniques. The distance for the raw image distribution is given as a baseline. For simplicity, we have used the Manhattan distance in our comparison. It can be seen that our method consistently works better than all existing methods currently used in the hardware assurance domain. The parameters for the filters were taken from earlier works \cite{c4, c14}. The results obtained from the performance analysis are also consistent with the results reported in these works. It can be observed from Table \ref{Results} that Curvature filter performs the worst with the majority of the required features being corrupted. Gaussian and Median filters performs comparably with noisy under/over-segmented but usable results. Anisometric Diffusion (AD) filter outperforms all of them. It can be attributed to the fact that AD requires an estimate of the noise in the image before processing. This prior information enables it to perform better. However, in case of samples with low noise like the metal layer, AD causes the metal features to merge making it unsuitable for RE. With the initial pre-processing in our approach, noise is considerably suppressed and a better estimate of the image histogram is obtained. This ensures consistent and effective segmentation of IC structures irrespective of the materials present in the SEM image.    

\section{Conclusion}

{}

In this paper, we were able to address and resolve a major problem in the area of IC RE, the segmentation of IC structures from poor quality SEM images. This was accomplished by extracting the histogram of the image, correcting it and segmenting the histogram based on the number of peaks in it. Even though this is the basic idea followed by most segmentation algorithms, our method has some unique advantages.

The algorithm strictly works on the histogram of the image. Hence, the size of the image does not effect the final segmentation. It does not try to model the noise sources or the features. Hence, expensive data collection sessions can be avoided. In addition, since the features are not modeled, partial occlusions of the features due to image stitching would not effect the segmentation process. The segmentation only relies on the working principles of the imaging modality to provide contrast between different constituent materials \cite{c12}.

Unlike some off-the-shelf segmentation algorithms, our method does not depend on the type of underlying distribution of the pixels belonging to each of the materials. Hence, all visible peaks are extracted from histogram by the algorithm even in presence of noise.

The algorithm does not require parameter fine-tuning. The only parameter used in the algorithm is the size of the smallest feature that needs to be extracted. This can be easily calculated from the raw image or set to the minimum size possible without effecting the final result. Finally, being unsupervised, the algorithm can be made completely automated with no human interaction.

The application of our algorithm on SEM images of IC indirectly reduces the time and labor required in imaging the die by enabling the use of fast acquired lower quality images. Considering the fact that imaging takes up most of the time in the RE process\cite{c10}, the algorithm would help in the complete RE of ICs, with ICs using both legacy and higher node technologies, in shorter time than previously possible.

\end{document}